\documentclass[twocolumn]{aastex62}

\usepackage{times}
\usepackage{graphicx}
\usepackage{epstopdf}
\usepackage{url}
\usepackage{tikz}
\usepackage{pgfplots}
\pgfplotsset{compat=newest}

\def\ergscm2 {erg\,s$^{-1}$cm$^{-2}$}

\def\cm2 {cm$^{-2}$}

\definecolor{niceblue}{RGB}{57,106,177}
\definecolor{nicered}{RGB}{204,37,41}

\definecolor{pointniceblue}{RGB}{57,106,177}
\definecolor{pointnicered}{RGB}{204,37,41}
\definecolor{pointnicegreen}{RGB}{62,150,81}
\definecolor{pointniceorange}{RGB}{218,124,48}

\shorttitle{\textsc{CRUSTAL FAILURES IN YOUNG MAGNETARS}}
\shortauthors{\textsc{Dehman et al.}}

\begin{document}
	
\title{ON THE RATE OF CRUSTAL FAILURES IN YOUNG MAGNETARS}

\correspondingauthor{Clara Dehman, email: dehman@ice.csic.es}

\author[0000-0003-0554-7286]{C. Dehman}
\affiliation{Institute of Space Sciences (ICE, CSIC), Campus UAB, Carrer de Can Magrans s/n, 08193, Barcelona, Spain}
\affiliation{Institut d'Estudis Espacials de Catalunya (IEEC), Carrer Gran Capit\`a 2--4, 08034 Barcelona, Spain}

\author[0000-0001-7795-6850]{D. Vigan\`o}
\affiliation{Institute of Space Sciences (ICE, CSIC), Campus UAB, Carrer de Can Magrans s/n, 08193, Barcelona, Spain}
\affiliation{Institut d'Estudis Espacials de Catalunya (IEEC), Carrer Gran Capit\`a 2--4, 08034 Barcelona, Spain}

\author[0000-0003-2177-6388]{N. Rea}
\affiliation{Institute of Space Sciences (ICE, CSIC), Campus UAB, Carrer de Can Magrans s/n, 08193, Barcelona, Spain}
\affiliation{Institut d'Estudis Espacials de Catalunya (IEEC), Carrer Gran Capit\`a 2--4, 08034 Barcelona, Spain}

\author[0000-0003-1018-8126]{J.A. Pons}
\affiliation{Departament de F\'{\i}sica Aplicada, Universitat d'Alacant, Ap. Correus 99, 03080 Alacant, Spain}

\author[0000-0002-3635-5677]{R. Perna}
\affiliation{Department of Physics and Astronomy, Stony Brook University, Stony Brook, NY 11794-3800, USA}
\affiliation{Center for Computational Astrophysics, Flatiron Institute, 162 5th Avenue, New York, NY 10010, \
USA}

\author[0000-0002-9575-6403]{A. Garcia-Garcia}
\affiliation{Institute of Space Sciences (ICE, CSIC), Campus UAB, Carrer de Can Magrans s/n, 08193, Barcelona, Spain}
\affiliation{Institut d'Estudis Espacials de Catalunya (IEEC), Carrer Gran Capit\`a 2--4, 08034 Barcelona, Spain}
\affiliation{Departament de F\'{\i}sica Aplicada, Universitat d'Alacant, Ap. Correus 99, 03080 Alacant, Spain}

\begin{abstract}

The activity of magnetars is powered by their intense and dynamic magnetic fields and has been proposed as the trigger to extragalactic Fast Radio Bursts. Here we estimate the frequency of crustal failures in young magnetars, by computing the magnetic stresses in detailed magneto-thermal simulations including Hall drift and Ohmic dissipation. The initial internal topology at birth is poorly known but is likely to be much more complex than a dipole. Thus, we explore a wide range of initial configurations, finding that the expected rate of crustal failures varies by orders of magnitude depending on the initial magnetic configuration. Our results show that this rate scales with the crustal magnetic energy, rather than with the often used surface value of the dipolar component related to the spin-down torque. The estimated frequency of crustal failures for a given dipolar component can vary by orders of magnitude for different initial conditions, depending on how much magnetic energy is distributed in the crustal non-dipolar components, likely dominant in newborn magnetars. The quantitative reliability of the expected event rate could be improved by a better treatment of the magnetic evolution in the core and the elastic/plastic crustal response, here not included. Regardless of that, our results are useful inputs in modelling the outburst rate of young Galactic magnetars, and their relation with the Fast Radio Bursts in our and other galaxies.
\end{abstract}


\section{Introduction}

Neutron stars (NSs) are the final remnants of many massive stars after their supernova explosions. They manifest themselves thanks to the huge magnetic energy stored inside and around them. The detected emission can be powered by the electromagnetic torque which induces rotational energy losses, by accretion, by crustal heating and by magnetic instabilities. In particular, magnetars, the most magnetic objects within the NS population, show magnetically-driven enhancements of their thermal and non-thermal emission, called outbursts \citep{beloborodov16,cotizelati18}. Their irregular and spectacular activity includes also short bursts and flares, believed to involve the magnetosphere, and possibly triggered by the interior dynamics.

Possibly related to magnetars \citep{margalit18} are Fast Radio Bursts (FRBs), very bright ($\sim$ 50 mJy - 100 Jy) millisecond-long bursts in the radio band, known for over a decade. Their distances as derived from the radio dispersion measures (DMs) has established their extragalactic nature, since the first identifications of their host galaxies confirmed their extragalactic nature  \citep[for recent reviews see, e.g.,][]{petroff19,cordes19}. Their short durations, coherent radio emission, and the discovery of several repeating FRBs \citep{spitler14,fonseca20}, strengthened their connection with  NSs, and in particular with strongly magnetized ones \citep{wadiasingh20,cheng20}.

Among the tens of theoretical models (see \citealt{platts19} for a list), \cite{lyutikov16} propose that FRBs are giant pulses coming from young NSs spinning at very fast rotation frequencies. On the other hand, many works assume that the star's magnetic energy is the ultimate engine behind the emission. For instance, \cite{lyubarsky14} propose maser synchrotron emission arising from far regions, where the magnetar wind interacts with the nebula inflated by the wind within the surrounding medium, and generates a forward and a reverse shock, both able to provide intense, coherent emission frequencies of tens to hundreds of MHz. A variant is the baryonic shell model, originally proposed by \citet{beloborodov17} and further developed by \citet{metzger19} and \citet{margalit20}, in which an ultra-relativistic flare collides with matter ejected from an earlier flare. \cite{kumar20} propose a model where a relatively small Alfv\'en disturbance at the surface, with amplitude $\delta B\sim 10^{11}$ G, is able to trigger the FRB emission via an efficient conversion to coherent curvature radiation from plasma bunches, accelerated by local electric fields.


The recent detection of a FRB-like radio bursts simultaneous with a bright hard X-ray burst from the Galatic magnetar SGR\,1935+2154 \citep{chime20,bochenek20,mereghetti20} showed for the first time that magnetar bursts can indeed be associated with FRB-like radio emission.
It has been speculated that the underlying mechanism is the same as for the extragalactic events, for which we see only the very bright events from a much larger volume \citep{margalit20}.

Several works, (e.g. \citealt{yuan20,lu20}) have proposed a magnetic disturbance from the surface driving the FRB. Regardless of the location and the exact mechanism, in the majority of magnetically-powered FRB models, the ultimate trigger to the event is expected to come from the interior of the NS. Internal magnetic fields evolve in the long term due to the Ohmic dissipation and Hall drift in the crust \citep{vigano13,pons19}. In the core, recent significant progress has been made in modeling and understanding the ambipolar diffusion and the evolving MHD equilibrium in superfluid (e.g. \cite{graber15,bransgrove17,gusakov19}) and non-superfluid cores \citep[e.g.,][]{castillo17,ofengeim18,dommes20,castillo20}, which for extreme values of magnetic field may proceed on relatively short timescales, thus changing the magnetic field on the crust-core boundary \citep{beloborodov16}.

In this Letter, we evaluate how often the crust of a magnetar will fail during the first centuries of its life, using an updated version of our magneto-thermal code \citep{vigano12}. This study improves and extends to the early stage previous results obtained for middle-aged magnetars \citep{perna11,pons11}, by including the Hall effect and up-to-date microphysics, but not the mechanisms in the core mentioned above. Since the early-stage dynamics are extremely sensitive to the initial conditions and the field topology, we perform the analysis under different assumptions on the initial magnetic field strength and geometry. We discuss our results in light of the current knowledge of FRB rates, recurrent times, and predictions in the FRB-magnetar models.

\section{Magnetic stresses from Magneto-thermal models}

During the first decades of a NS's life, the star quickly cools from its initial $10^{11}-10^{12}$ K to a few times $10^8$ K. As the temperature drops below the density-dependent melting value, the crust grows as a solid lattice of heavy ions. The crust freezing starts from the inner region a few minutes after birth, but takes several years to extend to the outer crust (e.g., Fig.~8 of \citealt{aguilera08b}). The crust is thought to provide an elastic response to stresses \citep{chugunov10} up to a maximum value:

\begin{equation}
    \sigma^{{\rm max}}= \bigg( 0.0195 - \frac{1.27}{\Gamma-71} \bigg)n_i\frac{Z^2 e^2}{a},
    \label{eq: sigma max}
\end{equation}

where $\Gamma=Z^2 e^2 / aT $ is the Coulomb coupling parameter, $a=\big[3/(4\pi n_i) \big]^{(1/3)}$ the ion sphere radius, $n_i$ the ion number density, $Z$ the charge number, $e$ the elementary charge, and $T$ the temperature. A recent calculation \citep{kozhberov20} confirms the validity of the approximation (Eq. \ref{eq: sigma max}).

As the magnetic field evolves, the magnetic tensor $M_{ij}(\vec{x},t)=B_i(\vec{x},t) B_j(\vec{x},t)/(4\pi)$ changes. Moreover, a stress quantified as a difference between $M_{ij}$ and an equilibrium $M_{ij}^{\rm eq}$, piles up in the crust. When it reaches the maximum stress locally allowed, $(M_{ij}-M_{ij}^{\rm eq})\simeq \sigma^{\rm max}(\vec{x})$, the crust fails. After the event, a new equilibrium state is established, the affected region freezes again and the crust starts to respond elastically, storing further stress until the cycle is repeated. 

An evaluation of the frequency of such events, which can potentially trigger magnetar activity, can be done by properly following the magneto-thermal evolution and the accumulated local stresses. Such simulations have been performed by \cite{perna11} and \cite{pons11}, considering the age range of the observed sources, $\sim 400-10^5$ yr. They used the magneto-thermal code by \cite{pons09}, not considering Hall effects. 

\begin{deluxetable*}{ccccccccccc}[t!]
\tablecaption{Initial models considered and events simulated: topology (where CT is a core-threaded configuration, CC is a crust-confined, P and T are the poloidal and toroidal field, D and Q mean purely dipolar and quadrupolar, and M a mixed multipolar configuration); surface value of the dipolar poloidal component $B_{\rm dip}$; magnetic energy $E_{\rm mag}^{\rm cr}$ stored in the crust, the fraction $E_{\rm tor}^{\rm cr}/E_{\rm mag}^{\rm cr}$ stored in the toroidal field; $N^x$ is the number of events shown during the first $x=$100, 400 and 1000 years; first event is the age at which the crust fails for the first time. Models marked with $^*$ show small local numerical instabilities, that can provide artificial events, quantifiable by up to a maximum of $30\%$ of the shown values $N$, thus not affecting the order of magnitude.
\label{tab: models}}
\tablecolumns{9}
\tablenum{1}
\tablewidth{0pt}
\tablehead{
\colhead{Model} &
\colhead{Topology} &
\colhead{$B_{\rm dip}$ } &
\colhead{$E_{\rm mag}^{\rm cr}$ } &
\colhead{$E_{\rm tor}^{\rm cr}/E_{\rm mag}^{\rm cr}$} &
\colhead{$N^{100}$} &
\colhead{$N^{400}$} &
\colhead{$N^{1000}$} &
\colhead{First event} \\
\colhead{} &
\colhead{} &
\colhead{[G]} & 
\colhead{[erg]} &
\colhead{} &
\colhead{} &
\colhead{} &
\colhead{} &
\colhead{[yr]}}
\startdata
\texttt{CrDip} & CC: PD+TQ & $10^{14}$ & $3.2\times 10^{46}$ & $\sim 50\%$ & 107 & 192 & 245 & 1.3 \\
\texttt{CrDipL} & CC: PD+TQ & $10^{13}$ & $3.2\times 10^{44}$ & $\sim 50\%$ & 0 & 0 & 0 & - \\
\texttt{CrDipH} & CC: PD+TQ & $3\times 10^{14}$ & $2.9 \times 10^{47}$ & $\sim 50\%$ & 2380 & 3315 & 3724 & 0.4 \\
\texttt{CrDipE} & CC: PD+TQ & $10^{15}$ & $3.2\times 10^{48}$ & $\sim 50\%$ & 23360 & 30842 & 34890 & 0.1 \\
\texttt{CrMultiL}$^*$ & CC: PM+TM & $10^{13}$ &  $9.3\times 10^{44}$ & $\sim 50\%$ &28 & 41 &  49 & 3.0 \\
\texttt{CrMultiH}$^*$ & CC: PM+TM & $10^{14}$ & $2.9\times 10^{47}$ & $\sim 50\%$ & 15538 & 21397 & 25513 & 0.1 \\
\texttt{CrMultiT}$^*$ & CC: PM+TM & $10^{13}$ & $4.8\times 10^{46}$  & $\sim 99 \%$ &525 &1335 & 1908  &0.7  \\
\texttt{CrDipD} & CC: PD+TD & $10^{14}$ & $3.4\times 10^{46}$ & $\sim 50\%$ & 88 & 176 & 254 & 1.6 \\
\texttt{CrDipP} & CC: PD+TQ & $10^{14}$ & $1.6\times 10^{46}$ & $\sim 1\%$ & 45 &  58 & 67 & 2.6 \\
\texttt{CoDipL} & CT: PD+TD & $10^{14}$ & $4.7\times 10^{44}$ & $\sim 50\%$ & 10 & 10 & 10 & 13.7 \\
\texttt{CoDipH}$^*$ & CT: PD+TD & $3 \times 10^{14}$ & $ 8.8 \times 10^{45}$ & $\sim 50\%$ & 160 & 168 & 174 &1.6  \\
\texttt{CoMulti} & CT: PM+TD & $  10^{14}$ & $ 7.5  \times 10^{46}$ & $ \sim 50 \%$ &1948  &  2265 &  2370 & 0.8 \\
\texttt{CoMultiP} & CT: PM+TD & $  10^{14}$ & $  2.6 \times 10^{46}$ & $\sim 1\%$ & 528 & 577 & 597 & 4.0 \\
\texttt{CoMultiPH}$^*$ & CT: PM+TD & $  3 \times 10^{14}$ & $   2.4 \times 10^{47}$ & $ \sim 1 \%$ & 9497 & 9997  & 10460  & 0.8 \\
\texttt{CoMultiPE}$^*$ & CT: PM+TD & $  6 \times 10^{14}$ & $   9.4 \times 10^{47}$ & $ \sim 1 \%$ &21090  & 23840  & 29065  & 0.4 \\
\enddata

\end{deluxetable*}

As in \cite{pons11} and \cite{perna11}, we assume that after a crustal failure event, the surrounding regions of the crust, being close to the maximum stress, $M_{ij}\geq \epsilon \sigma^{\rm max}$, are affected by the instability and resettle to equilibrium. In our simulations, we set the effective parameter to $\epsilon=0.9$. Note that increasing it to values closer to 1 will lead to more frequent events but smaller involved regions, and vice versa.

Our simulations present a few relevant simplifcations: (i) We do not include self-consistent magneto-elastic evolution of the crust \citep{cumming04,li16,thompson17,bransgrove17}, and the local magnetic field reconfiguration likely produced after a crustal failure; (ii) In the core we only consider the Ohmic term, which acts on timescales much longer than in the crust: in practice, the magnetic field does not evolve in the kyr here considered and a current sheets develops at the core surface; (iii) At the NS surface we impose a potential magnetic field solution (as in almost all studies, reviewed by \citealt{pons19}).


\begin{figure*}
\centering
\includegraphics[width=0.41\textwidth]{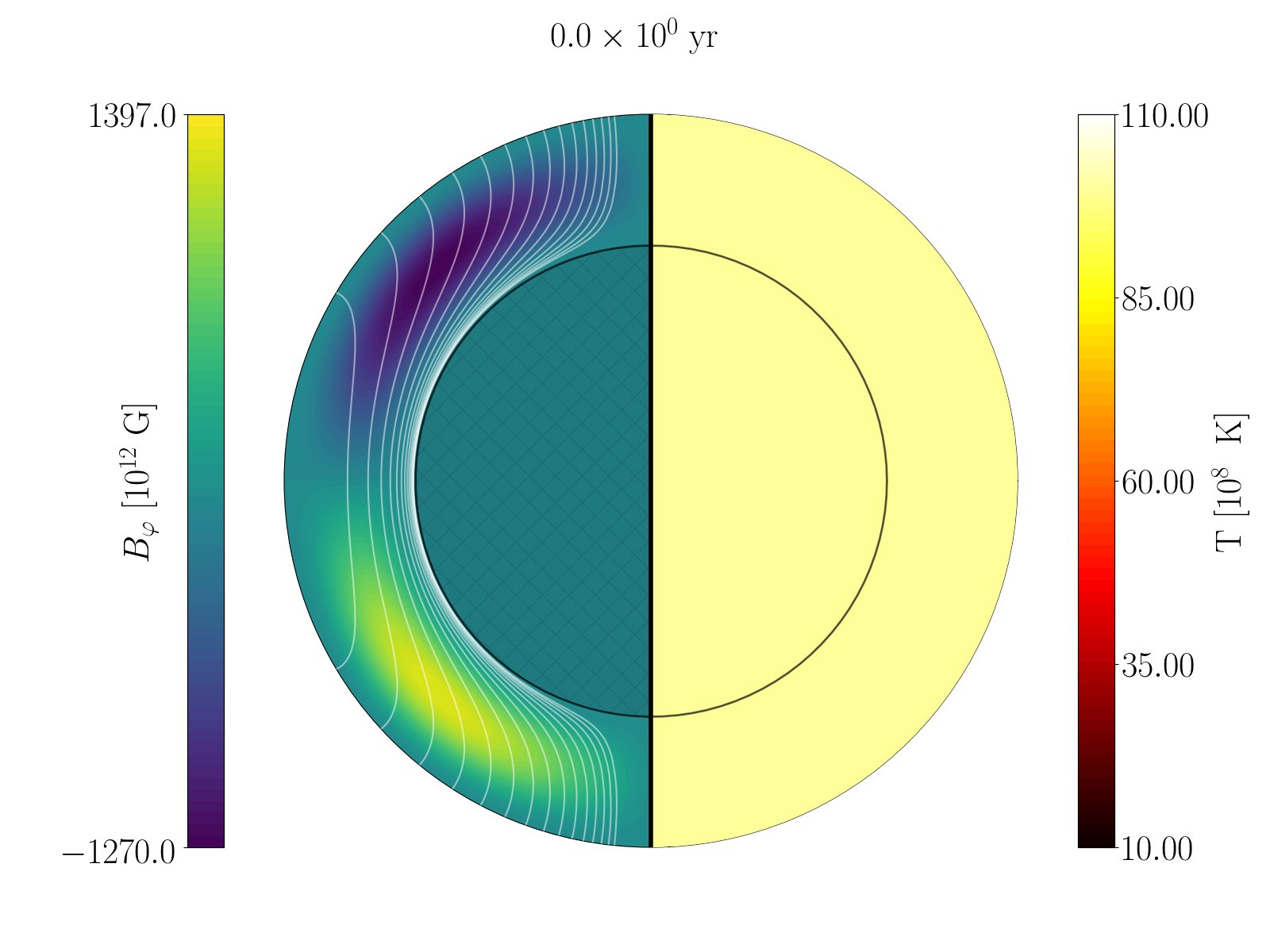}\hfill
\includegraphics[width=0.41\textwidth]{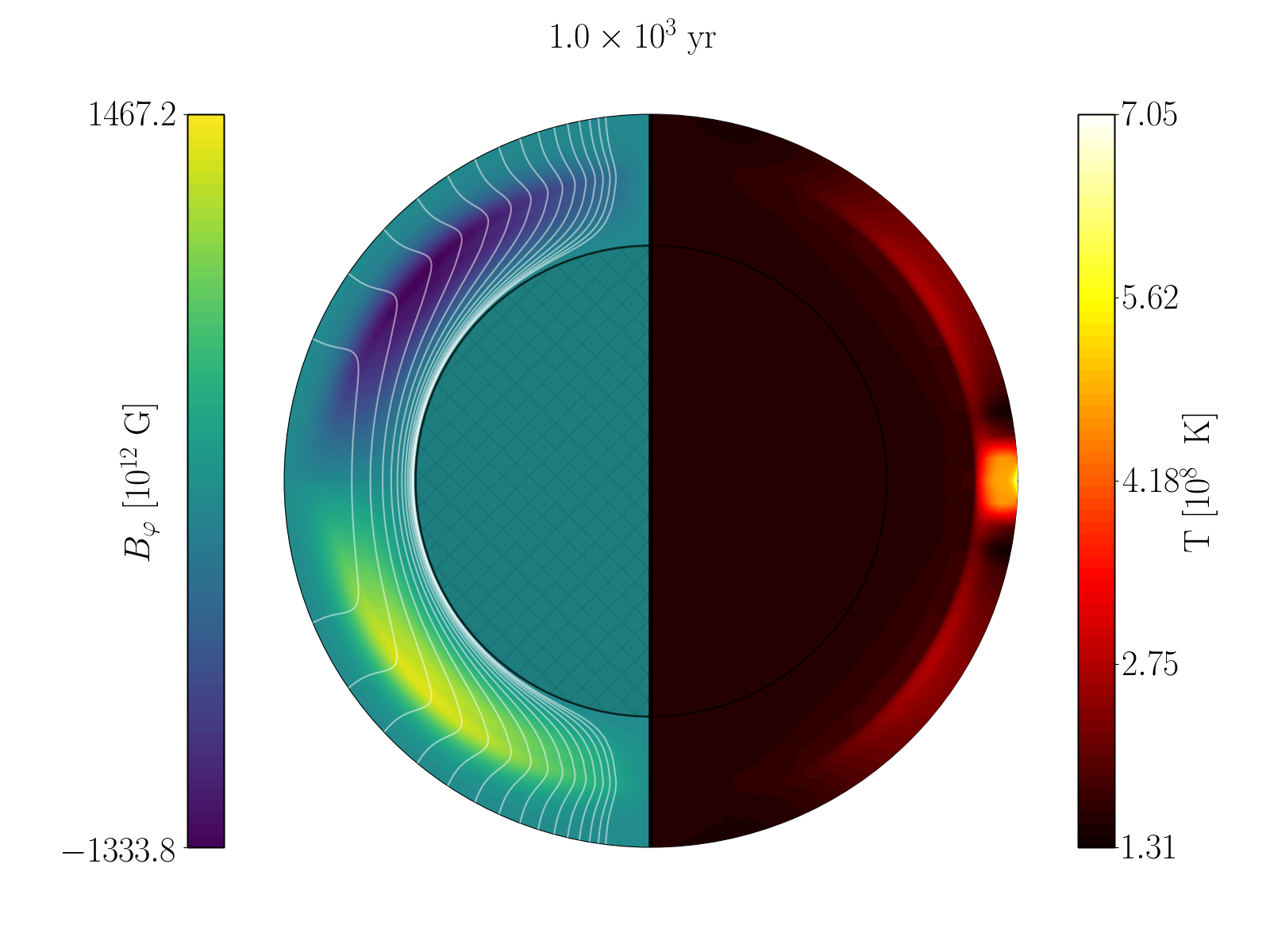}\vspace{0.1cm}
\includegraphics[width=0.41\textwidth]{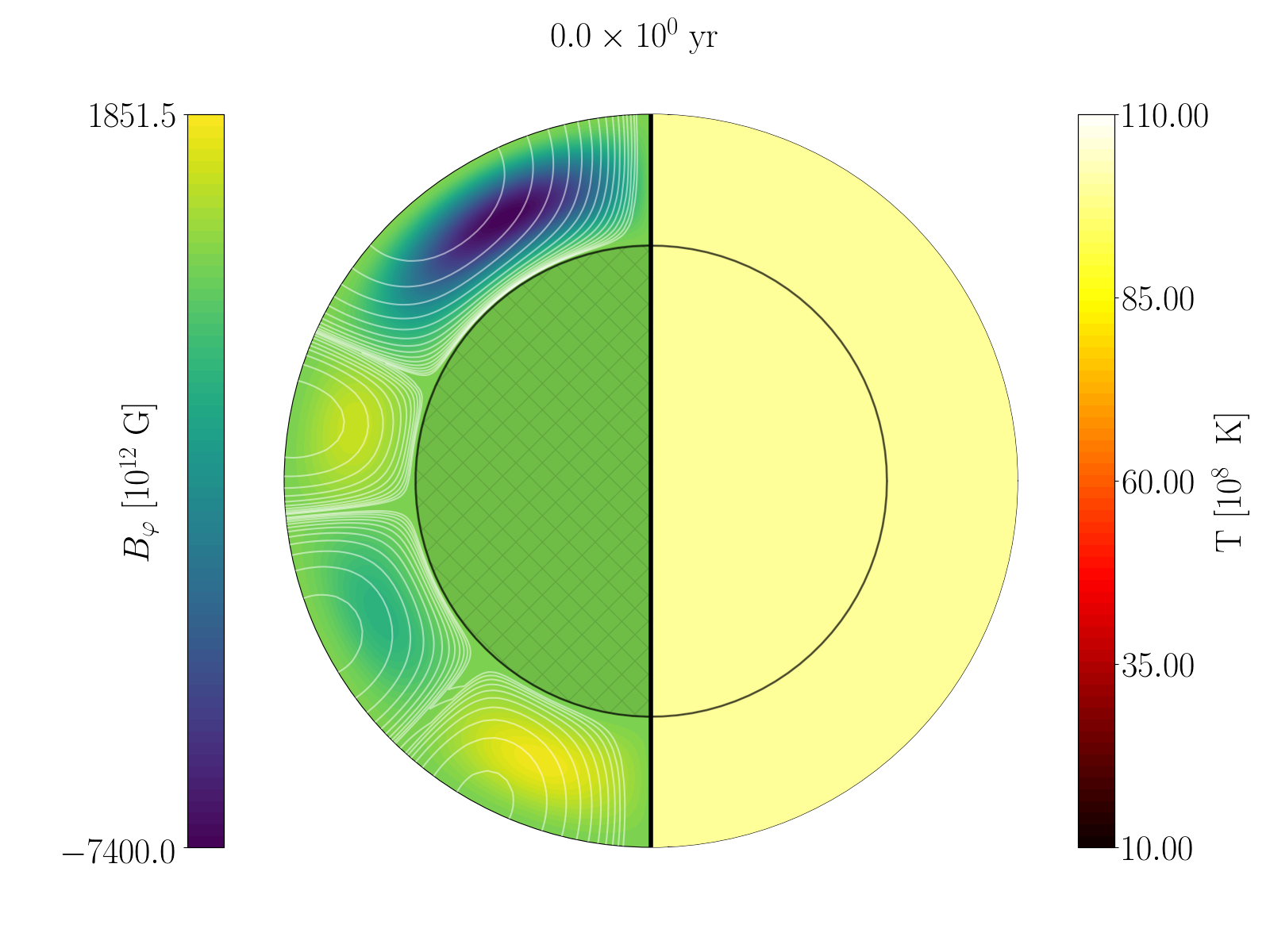}\hfill
\includegraphics[width=0.41\textwidth]{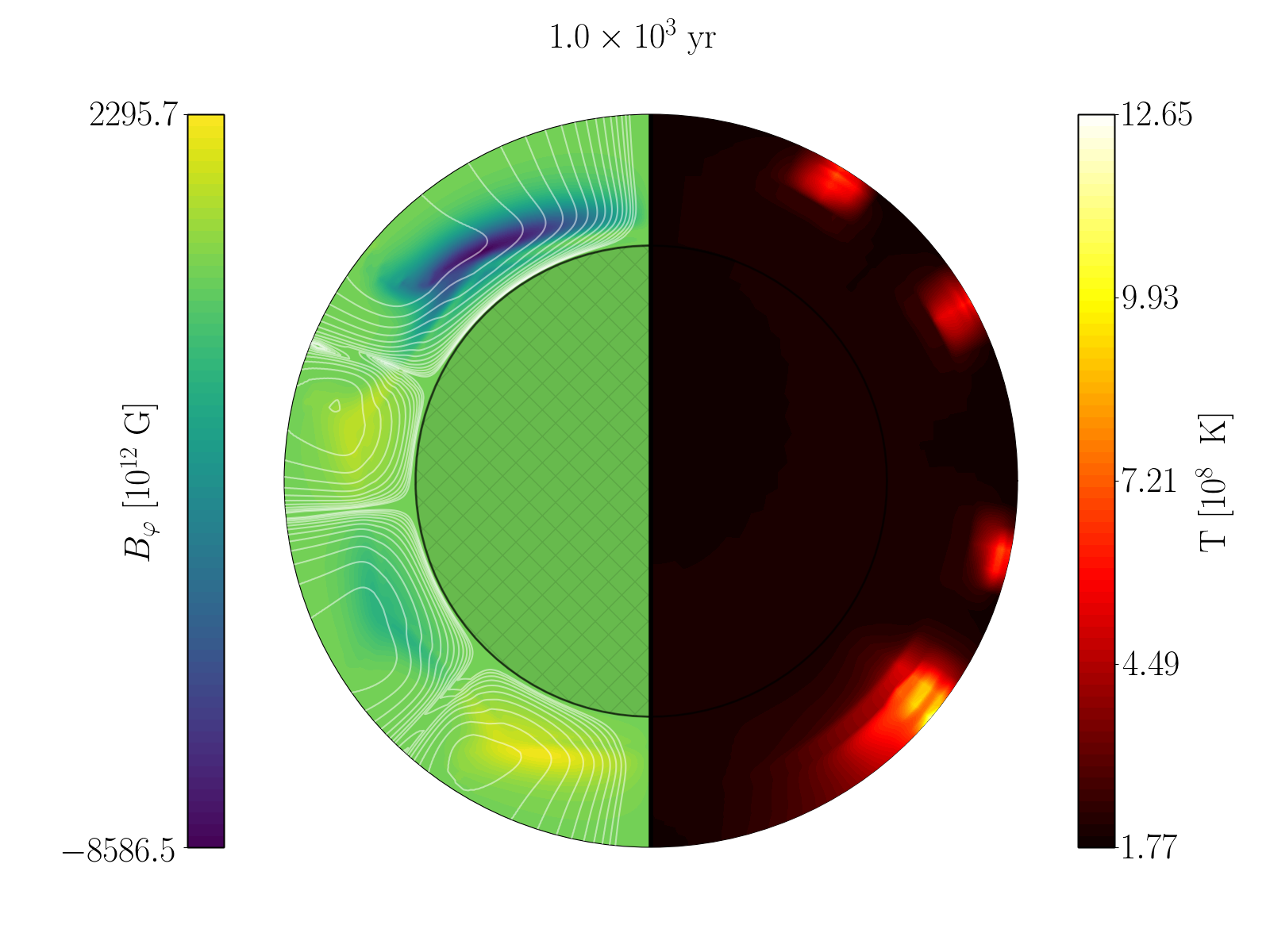}\vspace{0.1cm}
\includegraphics[width=0.41\textwidth]{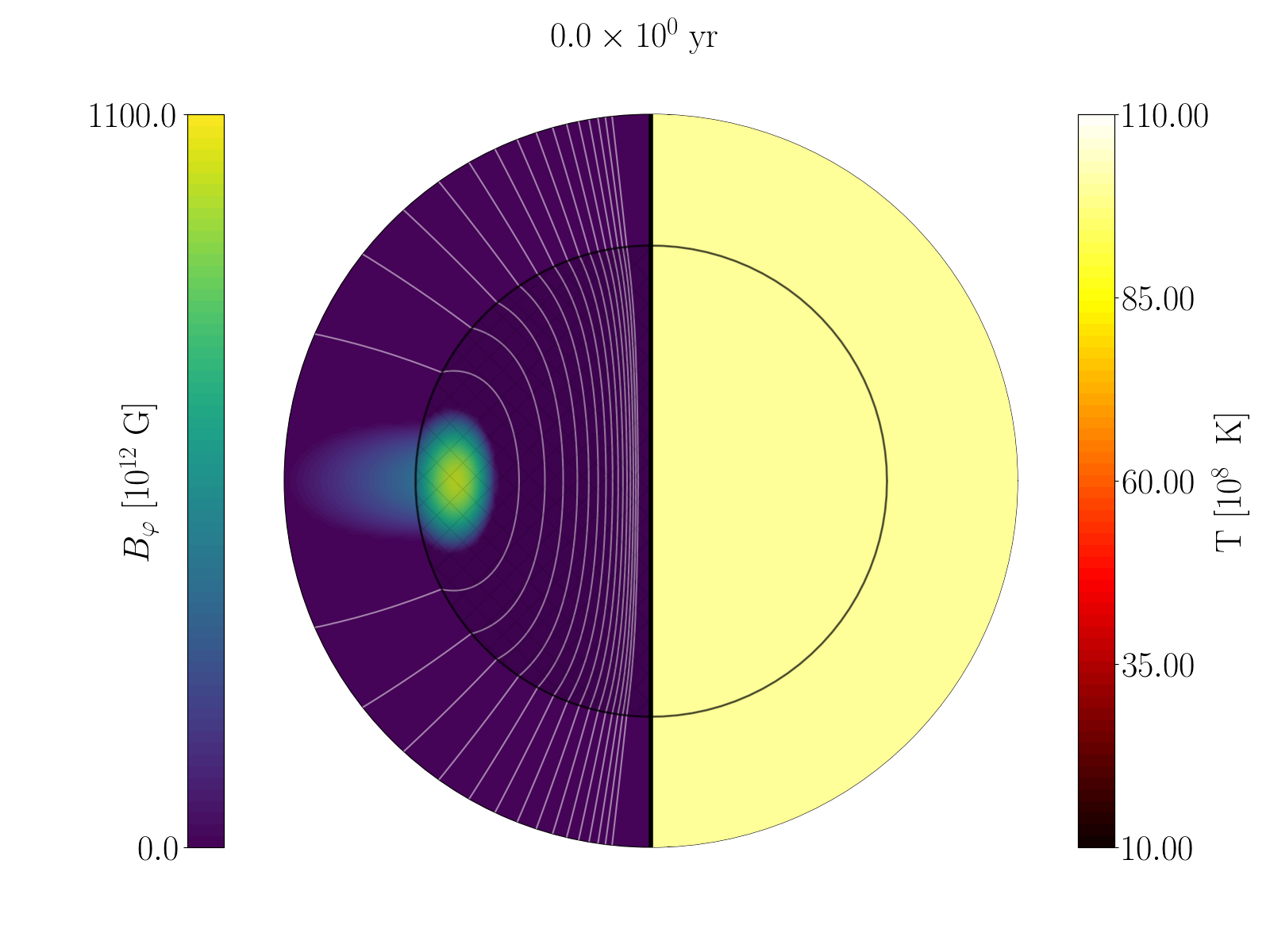}\hfill
\includegraphics[width=0.41\textwidth]{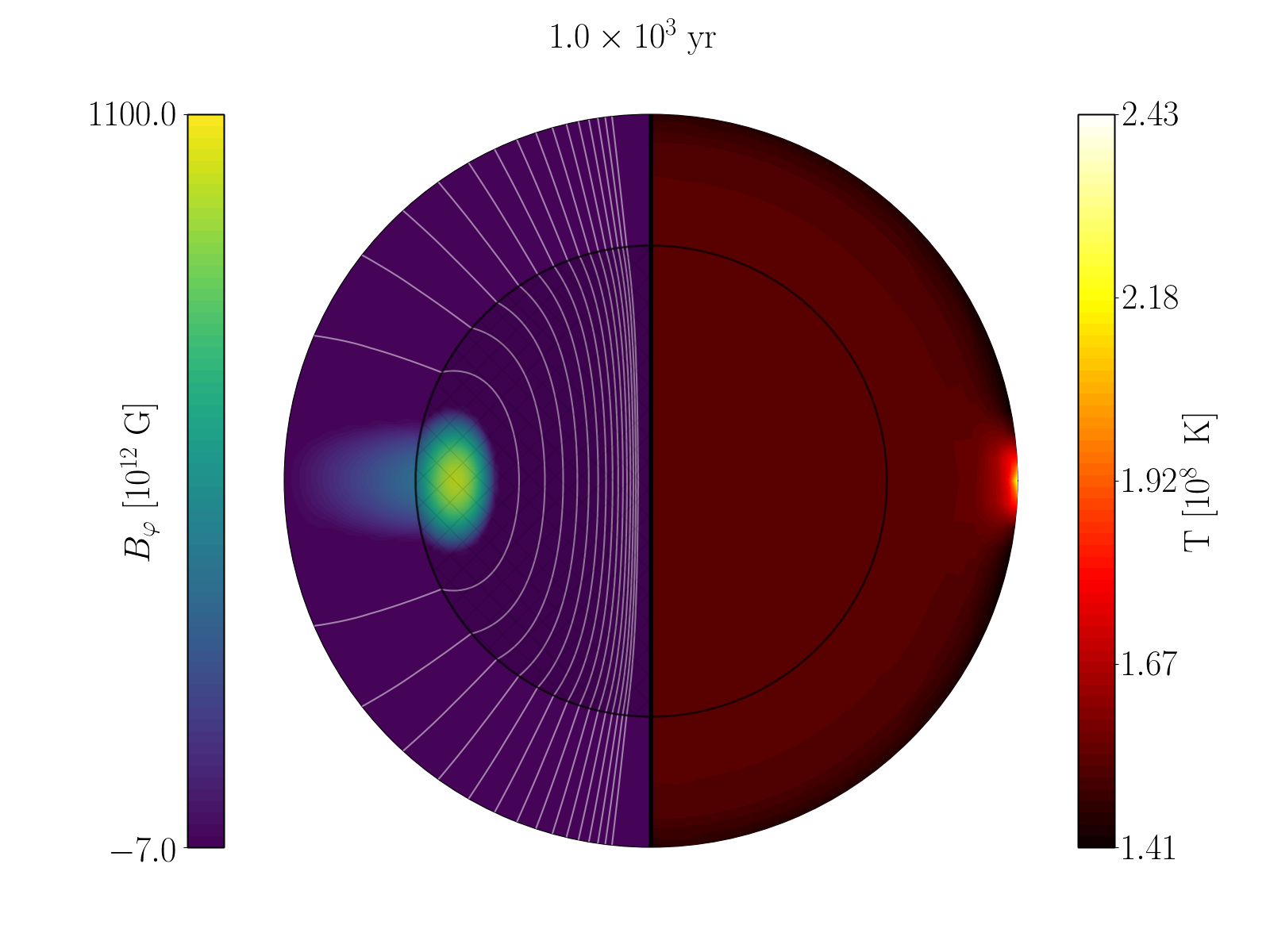}
\caption{Evolution of magnetic field and temperature for model {\tt CrDip} (top panels), the high field multipolar model {\tt CrMultiH} (central panels), and the low field core-threaded model {\tt CoDipL} (bottom panels), showing the meridional projection of the magnetic field lines (white lines) and the toroidal field (colors) on the left, and the internal temperature distribution (on the right), at the beginning (left panels) and after 1 kyr (right panels). The crust has been enlarged by a factor 8 for visualization purposes (thus only apparently bending the lines in the crust-core interface in the bottom panel). Movies of the crustal failure simulations for the different models in table \ref{tab: models} are illustrated in  \small \url{https://www.ice.csic.es/erc-magnesia/crustal-fractures-simulations/} }
\label{fig:2Dplots}
\end{figure*}


The main challenge in considering early ages is that the results are extremely sensitive to the initial magnetic field configuration, for both physical and numerical reasons.
Before freezing, the magnetized proto-neutron star fluid will settle to an MHD equilibrium/stationary state. Several works have proposed initial configurations, most of them based on a very large-scale and smooth initial magnetic field, usually with a purely dipolar poloidal component. However, nature typically provides much more turbulent, off-axis and non-symmetric configurations, as planetary and stellar magnetic fields show. Implementing such complicated configurations in simulations is a hard task. First of all, the possible solutions are potentially infinite and it is not clear which ones could be more realistic. Secondly, having small scales means having much faster Hall drift and whistler waves, characteristic of the Hall MHD equations which govern the solid crust. The propagation of such relatively fast transient waves dominates the dynamics and the piling up of the stresses over the first centuries. However, simulating such small-scale waves requires very fine space and time resolution, making the computational cost unfeasible, and the numerical instabilities become hardly controllable.

\section{Simulations}
\label{sec:runs}

\subsection{Initial configurations}

Aiming to reduce numerical instabilities which could provoke artificial crustal failures, we have used the latest version of the finite-volume axially-symmetric magneto-thermal code, improved in stability and efficiency compared to \cite{vigano12,vigano13}. It includes numerical methods which are suitable to the finite-volume integral version of the Hall induction equation, and are optimized in terms of stability, accuracy and efficiency. We also employ the most updated microphysical inputs \citep[for a review of cooling and transport see][]{potekhin15}.  
We make sure that results are close to the numerical convergence (we have variations of the event rate up to a maximum of $30\%$ for different resolutions, much less than the intra-model variability), using a spatial resolution of $100\times 200$ points in the angular and radial directions, and a fixed timestep of $\sim 10^{-4}-10^{-3} \hspace{0.05cm} $yr for the magnetic advance, and of $10^{-2}$ yr for the temperature evolution. 

To assess the sensitivity of results upon the uncertain initial conditions, we have considered very different topologies. Although not necessarily realistic, the variety considered allows us to explore the range of crustal failure events. They are summarized in Table \ref{tab: models}, together with the evolution of the events' frequency during the first 1 kyr of the NS's life.

\begin{figure*}[]
\includegraphics[width=.45\textwidth]{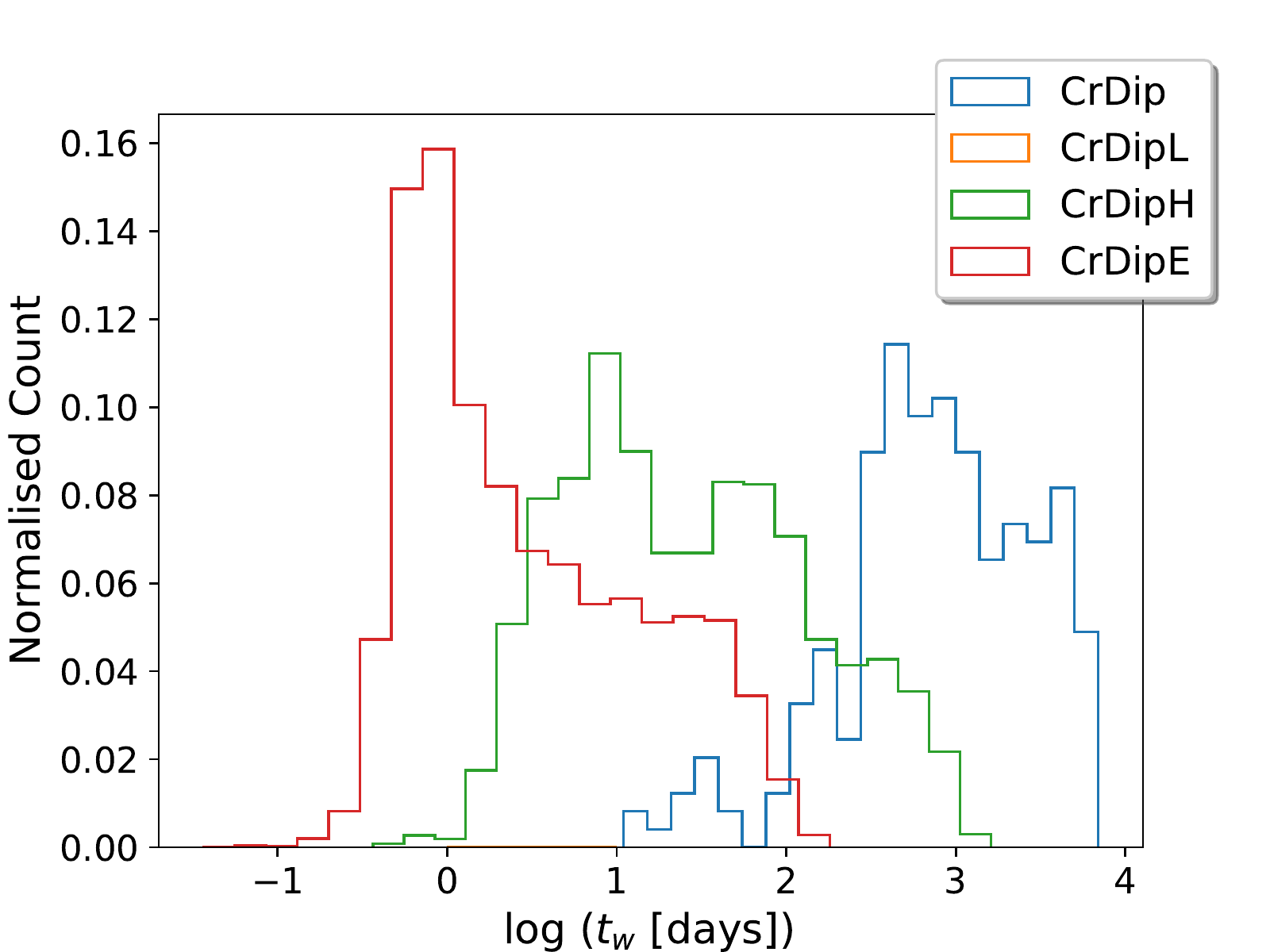}\hfill
\includegraphics[width=.45\textwidth]{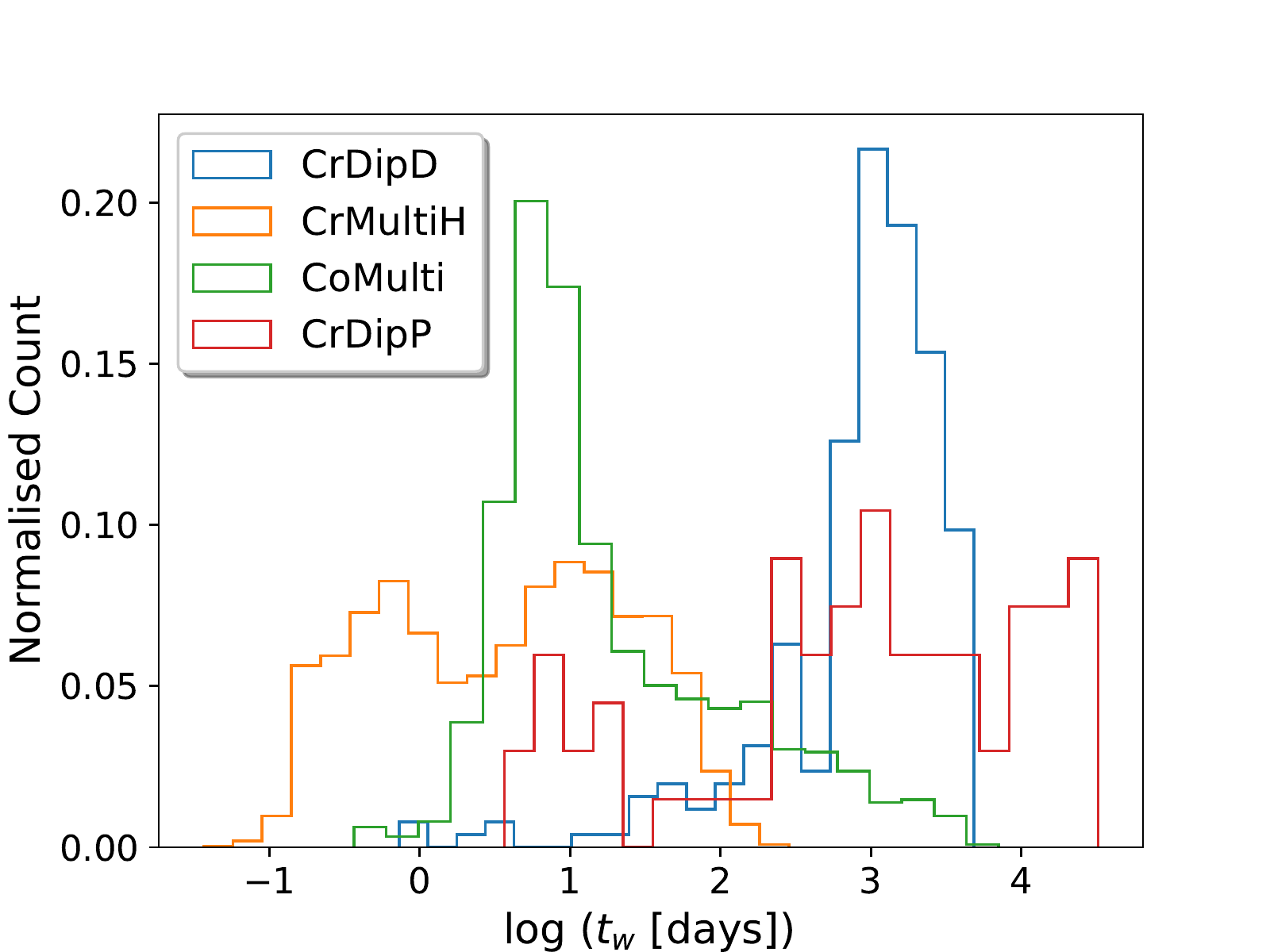}
\caption{Waiting time distributions during the first $1$ kyr of the NS's life. In the first panel, we investigate the effects of the magnetic strength on the crustal failures. In the second one, the study is focused on the field topology and on the impact of lowering the toroidal energy.}
\label{fig: WTdistributions crustal-confined}
\end{figure*} 

On one hand, we take into account crust-confined models with different weights in the initial multipoles. While it is unlikely that the field is completely expelled from the core, such a kind of topology can be useful to provide an upper limit on the frequency of outburst/burst activity triggered by the dynamics, since the field lines are not tied to the highly conductive core and the currents supporting the magnetic field flow entirely in the crust, where spatial and timescales are reduced compared to the core. Within these models, we vary the relative fraction of magnetic energy stored in different multipoles of both the poloidal and toroidal components, as follows:

\begin{itemize}
    \item In the fiducial model {\tt CrDip} we consider equipartition of magnetic energy between a dipolar poloidal magnetic field with polar surface value $B_{\rm dip}=10^{14}$~G and a toroidal quadrupolar field having the same magnetic energy as the poloidal one (and a maximum value $B_{\varphi}^{\rm max}=1.27\times 10^{15}$ G). 
    \item In the low, high, and extreme field models {\tt CrDipL}, {\tt CrDipH} and {\tt CrDipE}, we assume the same topology but with the magnetic field intensity rescaled by a factor 0.1, 3 and 10 respectively. 
    \item In the low and high field models {\tt CrMultiL} and {\tt CrMultiH}, we consider a topology with a similar contribution of the first four multipoles in both the poloidal and toroidal components, with an equipartition of magnetic energy between them. In {\tt CrMultiT}, the quadrupolar toroidal field accounts for $ 99\%$ of the total magnetic energy.
    \item In model {\tt CrDipD} a dipolar (instead of a quadrupolar) toroidal field is considered.
    \item In model {\tt CrDipP} we adopted a poloidal-dominated configuration, with the quadrupolar toroidal field accounting for less than $ 1\%$ of the total magnetic energy.
\end{itemize}

Additionally, we consider different core-threaded configurations:
\begin{itemize}
    \item Model {\tt CoDipL} (low field) and {\tt CoDipH} (high field), consisting of the often used large-scale twisted dipole plus torus configurations, as in \cite{akgun17}. Again, we set an equal fraction of poloidal and toroidal magnetic energy in the crust.
    \item In model {\tt CoMulti}, we consider a topology with similar contributions from the dipole and the quadrupole in the poloidal field, and a dipolar toroidal field, with an equipartition of magnetic energy between them.  On the other hand, for {\tt CoMultiP}, {\tt CoMultiPH}, and {\tt CoMultiPE}, the toroidal magnetic energy accounts for less than $ 1\%$ of the total magnetic energy, where {\tt H} and {\tt E} refer to high and extreme field models, respectively.
\end{itemize}

Note how models {\tt CrDip}, {\tt CrMultiH}, {\tt CrDipD}, {\tt CrDipP}, {\tt CoDipL}, {\tt CoMulti}, and {\tt CoMultiP} have the same dipolar polar magnetic field $B_{\rm dip}=10^{14} $ G, which rules the spin-down and, indirectly, the rotationally powered emission. However the magnetic energy stored in the crust, $E_{\rm mag}^{\rm cr}$, can vary by a few orders of magnitude, depending on the relative weight of the multipolar and toroidal components and on whether currents circulate only in the crust or not.

In Fig.~\ref{fig:2Dplots} we show the initial and evolved (at 1 kyr) magneto-thermal models {\tt CrDip} (top panels), \textbf{\tt CrMultiH} (central panels), and {\tt CoDipL} (bottom panels). The crustal-confined models show relevant magnetic activity (left hemispheres of each panel), with a local non-trivial rearrangement of the toroidal field distribution (colors) and a deformation of the poloidal field lines (white lines). The temperature distribution (right hemisphere) also evolves to be inhomogeneous, due to the transport anisotropy induced by the local intensity and direction of magnetic field.
On the other hand, the core-threaded model barely shows any evolution, due to the fact that the currents are mostly concentrated in the core, where the conductivity is very high and no Hall effect is present: the magnetic field lines are rooted and almost frozen in the core. The core-threaded cases give much slower dynamics than the crust-confined ones, due to the large characteristic spatial scales, and the fact that most of the supporting currents circulate in the highly conductive, non-solid core. A realistic configuration could be core-threaded, but consists of small scales, both in the core and the crust, so that the corresponding results should be effectively covered by our exploration of parameters. Note that, in all cases, the value of the dipolar component at the surface, $B_{\rm dip}$, decreases at most by a few percent.

\subsection{Event rate}

Focusing on the predicted event rate, the gross number for model {\tt CrDip} is $\sim 250$ events during the first 1 kyr of the NS's life, mostly concentrated in the first century. Furthermore, we emphasize that by changing the overall magnetic strength with respect to model {\tt CrDip}, the frequency decreases for the low-field dipolar model ({\tt CrDipL}), and increases for the high-field ({\tt CrDipE}) and the extreme field ({\tt CrDipH}) model, by $\sim 1$ and $\sim 2$ orders of magnitude, scaling the number of events as $N^{1000}\sim E_{\rm mag}^{\rm cr}/10^{44}$. The same scaling is noticed for the core-threaded models. There are no significant differences between a different topology for the toroidal field (quadrupolar in {\tt CrDip} and dipolar in {\tt CrDipD}), while important differences are noted if the magnetic energy is changed, maintaining the same $B_{\rm dip}$: poloidal-dominated {\tt CrDipP}, or considering multipolar configurations ({\tt CrMultiL}, {\tt CrMultiH}, and {\tt CrMultiT}).

For all models, the older the magnetar, the fewer the expected number of events. As shown in Table~\ref{tab: models}, $\sim 40-85\%$ of the events during the first 1000 years are concentrated in the first century, with a gradual decrease. This percentage decreases to $\sim 10-40 \%$ in the next 300 yrs, whereas in the last 600 years, the percentage oscillates between $\sim 5-20 \%$.  This is consistent with the findings for middle-age magnetars \citep{perna11,pons11}: magnetars become less and less active, but they can still show activity even at late stages.

The waiting time distributions are illustrated in Fig.~\ref{fig: WTdistributions crustal-confined}. In the first panel, we  compare models {\tt CrDip}, {\tt CrDipL}, {\tt CrDipH}, and {\tt CrDipE}. We find that increasing the strength of the magnetic field results in a shorter waiting time. This is also the case if we consider a multipolar topology (models {\tt CrMultiH} and {\tt CoMulti}), whereas lowering the fraction of the toroidal energy (model {\tt CrDipD}) has a tiny impact on the distribution. However, the waiting time is shifted from $\sim [10:10^4]$ to $ \sim [0.1:300]$ days, by increasing the field strength and considering the multipolar topology.

\begin{figure}
    \centering
        \includegraphics[width=.45\textwidth]{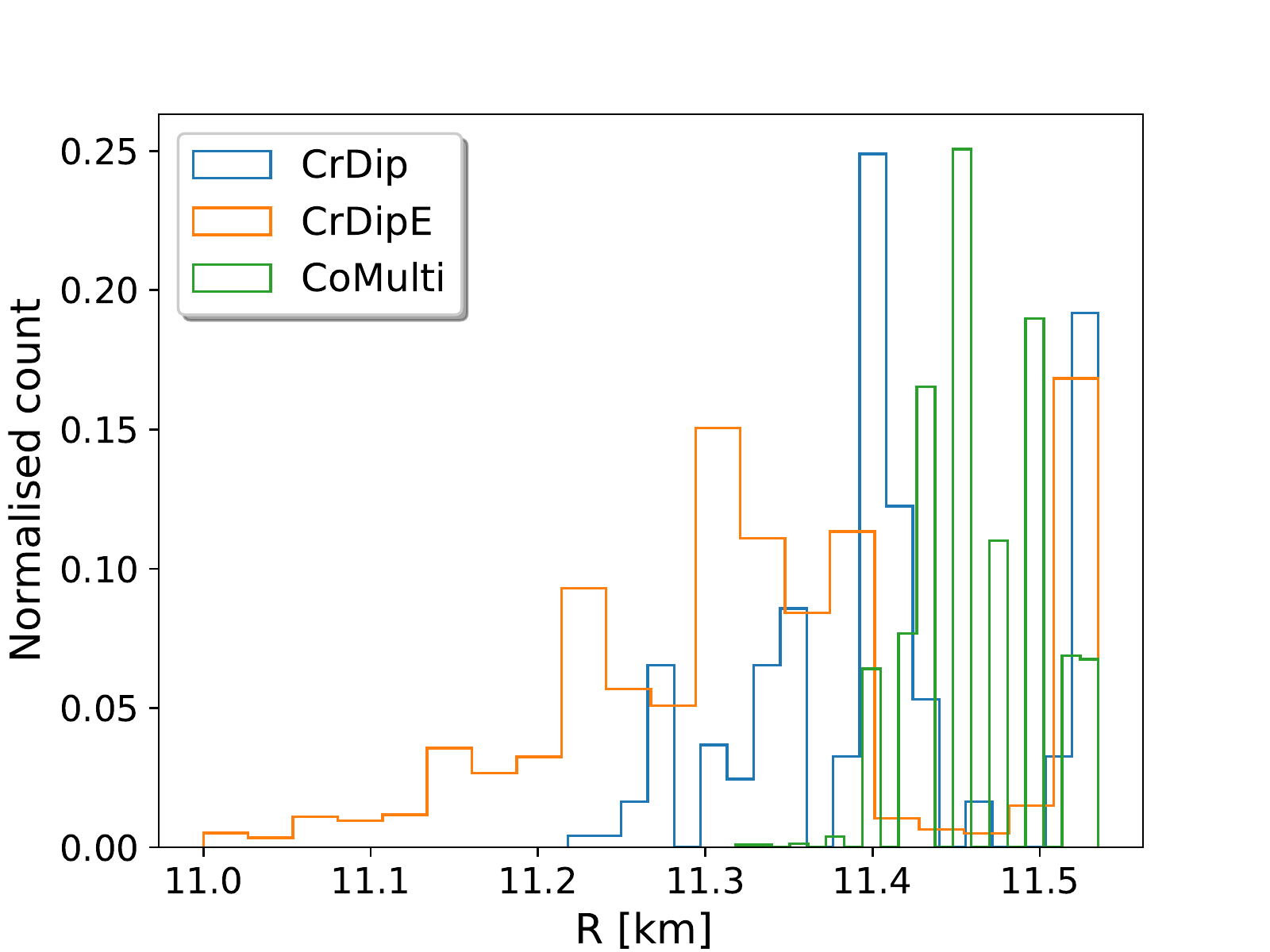}
    \caption{The location distribution of crustal failures as a function of the neutron star radius during the first 1 kyr of the NS's life. }
    \label{fig: radius distribution 025}
\end{figure}

In Fig.~\ref{fig: radius distribution 025}, we present the crustal failure distributions as a function of the NS radius for a set of models, i.e., model {\tt CrDip}, {\tt CrDipE}, and {\tt CoMulti}. Even though such distributions are partially model-dependent, most of the events happen in the outer crust, i.e., $R \sim [11.3: 11.55]$ km, because the maximum stress is smaller in the outermost, lighter layers. A non-negligible fraction of the events takes place in the outermost $50$ m of the NS, i.e., $\sim 25\%$.

\section{Discussion}
\label{sec:discussion}

Ideally, modeling the link between magnetospheric magnetar activity and crustal failure requires: (i) following the internal magnetic dynamics and how the magnetic stresses grow with time, (ii) understanding and quantifying the mechanisms leading to the propagation of the disturbance into the magnetosphere, (iii) describing how the magnetospheric instability/reconnection powers the emission.

This study focuses on the first issue, assessing the crustal failure frequency of newly born magnetars, under different initial topologies, through 2D magneto-thermal simulations, including the Hall term. Moreover, the presence of small-scale structures, very likely if the birth amplification involves turbulent processes, can increase the expected event rate, through initial enhanced Hall dynamics. More factors can further extend the range of events found, among which we must mention the magnetic field boundary conditions (here considered as potential, possibly inadequate for the dense environment around a newborn magnetar), the fixed effective parameter $\epsilon$, controlling how large is the region undergoing failure at the same time.

\begin{figure}[!t]
  \begin{tikzpicture}
    \begin{axis}[
      xlabel={$E^{\rm cr}_{\rm mag} [erg]$},
      ylabel={$N^{1000}$},
      height=6.5cm,
      width=0.95\linewidth,
      grid=major,
      enlargelimits=false,
      yticklabel style={
        /pgf/number format/fixed,
      },
      xmode=log,
      ymode=log,
      ymin=5,
      ymax=5e4,
      xmin=1e44,
      xmax=1e49,
      legend pos=north west,
      scaled ticks=false]
      \addplot[
            scatter,%
            scatter/@pre marker code/.code={%
                \edef\temp{\noexpand\definecolor{mapped color}{rgb}{\pgfplotspointmeta}}%
                \temp
                \scope[draw=mapped color!80!black,fill=mapped color]%
            },
            scatter/@post marker code/.code={%
                \endscope
            },
            only marks,
            mark=*,
            point meta={TeX code symbolic={
                \edef\pgfplotspointmeta{\thisrow{RED},\thisrow{GREEN},\thisrow{BLUE}}%
            }},
      ] table [x=Emag,y=N1000]{./frb.txt};
      \addlegendimage{only marks,mark=*, color=pointniceblue}
      \addlegendimage{only marks,mark=*, color=pointnicegreen}
      \addlegendimage{only marks,mark=*, color=pointniceorange}
      \addlegendimage{only marks,mark=*}
      \addlegendimage{only marks,mark=*, color=pointnicered}
      \legend{,$B_{\rm dip} = 10^{13}$, $B_{\rm dip} = 10^{14}$, $B_{\rm dip} = 3\times 10^{14}$,$B_{\rm dip} = 6\times 10^{14}$, $B_{\rm dip} = 10^{15}$}
\end{axis}
  \end{tikzpicture}
  \caption{Number of events in the first kyr as a function of the crustal magnetic energy. Each point represents a specific model, with colors indicating the value of $B_{\rm dip}$ as in the legend. } 
  \label{fig:events}
\end{figure}
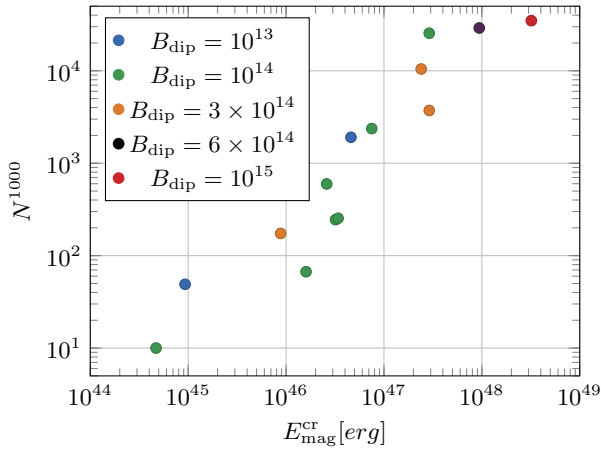

The main finding is that the event frequency scales linearly with the magnetic energy stored in the crust, with some intrinsic dispersion given by the topology, $E_{\rm mag}^{\rm cr}$ (see Fig.~\ref{fig:events}), which strongly depends on the assumed and highly uncertain field topology.

Given the simplifications in our approach, additional words of caution are needed in the interpretation of the results:\\
(i) We do not include the recent advances in the modeling of the fluid motion in NS core (with or without superfluidity), limiting to either confine the field to the crust, or to consider only the very slow Ohmic dissipation. In particular, the ambipolar diffusion could drive (slightly) faster changes and higher rates \citep{beloborodov16}.\\
(ii) Magneto-elastic simulations, and the inclusion of the local field rerrangement and dissipation, could reduce the rates, especially the most extreme ones (thus potentially weakening the linear trend in the upper range). Moreover, the critically stressed crust may behave plastically \citep{kobyakov14}, leading to Hall drift and thermoplastic waves \citep{beloborodov14,li16} and partially hampering the Hall dynamics \citep{lander19}.\\
(iii) 3D dynamics and more realistic boundary conditions could also have a quantitative impact, difficult to guess a-priori.

We expect that overcoming such caveats above will systematically change the quantitative results, but not the important conclusion that the dipolar field strength is a barely relevant parameter when considering a young magnetar's activity. As a matter of fact, for a given $B_{\rm dip}$, our calculations show a very broad range of crustal event rates. Even a relatively modest $B_{\rm dip}=10^{14}$ G could potentially provide almost no events (large-scale field penetrating in the core) or $N \sim 1$ event per day, if most of the magnetic energy is hidden in the form of crustal multipolar and toroidal fields. The expected crustal failure rate is, thus, strongly dependent on the initial topology of the magnetic field. These results shed light on a potential issue of most FRB-magnetars models, which are usually focused on the value of $B_{\rm dip}$ only.


\subsection{Comparison with Fast Radio Bursts and Galactic magnetars}

Our calculations have implications both for FRBs (if indeed associated with magnetars), as well as for Galactic magnetars. 
In both scenarios, the observed event rate can be summarized as
\begin{equation}
N_{\rm obs}=N_{\rm failure} \times N_{\rm sources} \times x_{\rm vis}\,,
\end{equation}
where $N_{\rm sources}$ is the number of potential sources (young magnetars within a certain observable volume), $N_{\rm failure}$ is the crustal event rate per star calculated in our study, and $x_{\rm vis}$ is the effective fraction of events that will actually be detected. The latter number is basically unconstrained a-priori, depending on the specific mechanisms ultimately leading to a detection, to observational biases, and to both physical (beaming, distance, absorption, etc.) and instrumental (sensitivity, field of view, band-width, etc.) limits.

In the case of magnetar bursting activity happening in the magnetosphere, $x_{\rm vis}$ depends on the crustal response to failure (one can expect that only the outermost events can further affect the magnetosphere), the propagation of the disturbance outside, the magnetospheric dynamics and the emission mechanism. Besides these fundamental physical uncertainties, emission beaming, distance and absorption will affect $x_{\rm vis}$, resulting in an expected value much smaller than one.

For FRBs, according to \cite{rane16}, the observed rate is estimated to be $N_{\rm obs} \sim 10^3–10^4$ per day per sky above 4 mJy per ms. Considering as a reference the number of galaxies within the distance of the prototypical FRB repeater FRB121112, $\sim 10^7$, and assuming  $\sim 10$ magnetar per galaxy with $\lesssim $ 1000 years, we would expect a potential number of FRB-emitting magnetars of about $N_{\rm sources} \sim 10^8$. Equating these numbers, we obtain a constraint $x_{\rm vis}^{\rm frb}\times N_{\rm failure} \sim 10^{-5}-10^{-4}$ per day. This means that models with high crustal magnetic energy, $E_{\rm mag}^{\rm cr}\gtrsim 10^{46}$ erg, which have $N_{\rm failure}\sim 1$ per day, are compatible with $x_{\rm vis} \ll 1$.

Similarly, very young ($\lesssim 1$ kyr) Galactic magnetars are expected to be $N_{\rm sources} \sim 10$. The observed rate of outbursts for them (excluding older magnetars) is $N_{\rm obs} \sim 10^{-3}-10^{-4}$ per day \citep{cotizelati18}. Therefore, $x_{\rm vis}^{\rm galactic}\times N_{\rm failure} \sim 10^{-5}-10^{-4}$, similarly to the FRBs.

This preliminary estimate, although plagued by uncertainties on the total number of sources and the still incomplete statistics, shows that the range of events detected per star, $x_{\rm vis}\times N_{\rm failure}$, in the FRB and Galactic magnetar outburst scenarios, are at least compatible. This points to a possible common origin of crustal triggers, even though they manifest in different ways.

\section{Conclusions}

Our simulations show that the dipolar component, on which most FRB-magnetar models are currently based, plays a minor role in determining the bursting activity of the magnetosphere if triggered by crustal failures. Providing reliable numbers for the expected event rate is intrinsically difficult, mainly for the intrinsic limitations of the approach and the highly uncertain initial configuration of the magnetic field. Here we have shown how, by varying the configuration from a large-scale, smooth core-threaded field, to a multipolar field confined in the crust, the calculated crustal failure rate changes by several orders of magnitude, with the same dipolar poloidal value, $B_{\rm dip}$. However, we have found that the crustal magnetic energy is instead a good tracer of the expected number of events, regardless of the exact topology.

Further observations and a better understanding of the mechanisms leading to the propagation of the disturbance and the emission of bursts, will help constrain the fraction of events that can be detectable ($x_{\rm vis}$). If this parameter was constrained even at an order-of-magnitude level, it would allow us to constrain the initial topology of the magnetic field. 



\acknowledgments
We thank Alice Borghese, Francesco Coti Zelati, and Vanessa Graber for helpful suggestions. CD, DV, NR and AGG are supported by the ERC Consolidator Grant ``MAGNESIA" (nr.817661) and acknowledge funding from grants SGR2017-1383 and PGC2018-095512-BI00. JAP acknowledges support by the Generalitat Valenciana (PROMETEO/2019/071) and by AEI grant PGC2018-095984-B-I00. RP acknowledges support from NSF award AST-1616157. We acknowledge support from the PHAROS COST Action (CA16214).

\bibliographystyle{aasjournal}

\begin{thebibliography}{}
\expandafter\ifx\csname natexlab\endcsname\relax\def\natexlab#1{#1}\fi
\providecommand{\url}[1]{\href{#1}{#1}}
\providecommand{\dodoi}[1]{doi:~\href{http://doi.org/#1}{\nolinkurl{#1}}}
\providecommand{\doeprint}[1]{\href{http://ascl.net/#1}{\nolinkurl{http://ascl.net/#1}}}
\providecommand{\doarXiv}[1]{\href{https://arxiv.org/abs/#1}{\nolinkurl{https://arxiv.org/abs/#1}}}

\bibitem[Aguilera et al.(2008)]{aguilera08b} Aguilera, D.~N., Pons, J.~A., \& Miralles, J.~A.\ 2008, \aap, 486, 255

\bibitem[Akg{\"u}n et al.(2017)]{akgun17} Akg{\"u}n, T., Cerd{\'a}-Dur{\'a}n, P., Miralles, J.~A., et al.\ 2017, \mnras, 472, 3914

\bibitem[Beloborodov \& Levin(2014)]{beloborodov14} Beloborodov, A.~M. \& Levin, Y.\ 2014, \apjl, 794, L24

\bibitem[Beloborodov \& Li(2016)]{beloborodov16} Beloborodov, A.~M. \& Li, X.\ 2016, \apj, 833, 261

\bibitem[Beloborodov(2017)]{beloborodov17} Beloborodov, A.~M.\ 2017, \apjl, 843, L26

\bibitem[Bochenek et al.(2020)]{bochenek20} Bochenek, C.~D., Ravi, V., Belov, K.~V., et al.\ 2020, arXiv:2005.10828

\bibitem[Bransgrove et al.(2018)]{bransgrove17} Bransgrove, A., Levin, Y., \& Beloborodov, A.\ 2018, \mnras, 473, 2771

\bibitem[Castillo et al.(2017)]{castillo17} Castillo, F., Reisenegger, A., \& Valdivia, J.~A.\ 2017, \mnras, 471, 507

\bibitem[Castillo et al.(2020)]{castillo20} Castillo, F., Reisenegger, A., \& Valdivia, J.~A.\ 2020, \mnras, doi:10.1093/mnras/staa2543

\bibitem[Cheng et al.(2020)]{cheng20} Cheng, Y., Zhang, G.~Q., \& Wang, F.~Y.\ 2020, \mnras, 491, 1498

\bibitem[The CHIME/FRB Collaboration et al.(2020)]{chime20} The CHIME/FRB Collaboration, :, Andersen, B.~C., et al.\ 2020, arXiv:2005.10324

\bibitem[Chugunov \& Horowitz(2010)]{chugunov10} Chugunov, A.~I. \& Horowitz, C.~J.\ 2010, \mnras, 407, L54

\bibitem[Cordes \& Chatterjee(2019)]{cordes19} Cordes, J.~M. \& Chatterjee, S.\ 2019, \araa, 57, 417

\bibitem[Coti Zelati et al.(2018)]{cotizelati18} Coti Zelati, F., Rea, N., Pons, J.~A., et al.\ 2018, \mnras, 474, 961

\bibitem[Cumming et al.(2004)]{cumming04} Cumming, A., Arras, P., \& Zweibel, E.\ 2004, \apj, 609, 999

\bibitem[Dommes et al.(2020)]{dommes20} Dommes, V.~A., Gusakov, M.~E., \& Shternin, P.~S.\ 2020, \prd, 101, 103020

\bibitem[Fonseca et al.(2020)]{fonseca20} Fonseca, E., Andersen, B.~C., Bhardwaj, M., et al.\ 2020, \apjl, 891, L6

\bibitem[Graber et al.(2015)]{graber15} Graber, V., Andersson, N., Glampedakis, K., et al.\ 2015, \mnras, 453, 671

\bibitem[Gusakov(2019)]{gusakov19} Gusakov, M.~E.\ 2019, \mnras, 485, 4936

\bibitem[Horowitz \& Hughto(2008)]{horowitz08} Horowitz, C.~J. \& Hughto, J.\ 2008, arXiv:0812.2650

\bibitem[Kobyakov \& Pethick(2014)]{kobyakov14} Kobyakov, D. \& Pethick, C.~J.\ 2014, \prl, 112, 112504

\bibitem[Kozhberov \& Yakovlev(2020)]{kozhberov20} Kozhberov, A.~A. \& Yakovlev, D.~G.\ 2020, \mnras, doi:10.1093/mnras/staa2715

\bibitem[Kumar \& Bo{\v{s}}njak(2020)]{kumar20} Kumar, P. \& Bo{\v{s}}njak, {\v{Z}}.\ 2020, \mnras, 494, 2385

\bibitem[Lander et al.(2015)]{lander15} Lander, S.~K., Andersson, N., Antonopoulou, D., et al.\ 2015, \mnras, 449, 2047

\bibitem[Lander \& Gourgouliatos(2019)]{lander19} Lander, S.~K. \& Gourgouliatos, K.~N.\ 2019, \mnras, 486, 4130

\bibitem[Li et al.(2016)]{li16} Li, X., Levin, Y., \& Beloborodov, A.~M.\ 2016, \apj, 833, 189

\bibitem[Lu et al.(2020)]{lu20} Lu, W., Kumar, P., \& Zhang, B.\ 2020, \mnras, 498, 1397

\bibitem[Lyubarsky(2014)]{lyubarsky14} Lyubarsky, Y.\ 2014, \mnras, 442, L9

\bibitem[Lyutikov et al.(2016)]{lyutikov16} Lyutikov, M., Burzawa, L., \& Popov, S.~B.\ 2016, \mnras, 462, 941

\bibitem[Marcote et al.(2020)]{marcote20} Marcote, B., Nimmo, K., Hessels, J.~W.~T., et al.\ 2020, \nat, 577, 190

\bibitem[Margalit \& Metzger(2018)]{margalit18} Margalit, B. \& Metzger, B.~D.\ 2018, \apjl, 868, L4

\bibitem[Margalit et al.(2020)]{margalit20} Margalit, B., Beniamini, P., Sridhar, N., et al.\ 2020, \apjl, 899, L27

\bibitem[Mereghetti et al.(2020)]{mereghetti20} Mereghetti, S., Savchenko, V., Ferrigno, C., et al.\ 2020, \apjl, 898, L29

\bibitem[Metzger et al.(2019)]{metzger19} Metzger, B.~D., Margalit, B., \& Sironi, L.\ 2019, \mnras, 485, 4091

\bibitem[Michilli et al.(2018)]{michilli18} Michilli, D., Seymour, A., Hessels, J.~W.~T., et al.\ 2018, \nat, 553, 182

\bibitem[Ofengeim \& Gusakov(2018)]{ofengeim18} Ofengeim, D.~D. \& Gusakov, M.~E.\ 2018, \prd, 98, 043007

\bibitem[Passamonti et al.(2017)]{passamonti17} Passamonti, A., Akg{\"u}n, T., Pons, J.~A., et al.\ 2017, \mnras, 469, 4979

\bibitem[Perna \& Pons(2011)]{perna11} Perna, R. \& Pons, J.~A.\ 2011, \apjl, 727, L51

\bibitem[Petroff et al.(2019)]{petroff19} Petroff, E., Hessels, J.~W.~T., \& Lorimer, D.~R.\ 2019, \aapr, 27, 4

\bibitem[Platts et al.(2019)]{platts19} Platts, E., Weltman, A., Walters, A., et al.\ 2019, \physrep, 821, 1

\bibitem[Pons et al.(2009)]{pons09} Pons, J.~A., Miralles, J.~A., \& Geppert, U.\ 2009, \aap, 496, 207

\bibitem[Pons \& Perna(2011)]{pons11} Pons, J.~A. \& Perna, R.\ 2011, \apj, 741, 123

\bibitem[Pons \& Vigan{\`o}(2019)]{pons19} Pons, J.~A. \& Vigan{\`o}, D.\ 2019, Living Reviews in Computational Astrophysics, 5, 3

\bibitem[Potekhin et al.(2015)]{potekhin15} Potekhin, A.~Y., Pons, J.~A., \& Page, D.\ 2015, \ssr, 191, 239

\bibitem[Rane et al.(2016)]{rane16} Rane, A., Lorimer, D.~R., Bates, S.~D., et al.\ 2016, \mnras, 455, 2207

\bibitem[Spitler et al.(2014)]{spitler14} Spitler, L.~G., Cordes, J.~M., Hessels, J.~W.~T., et al.\ 2014, \apj, 790, 101

\bibitem[Suvorov \& Kokkotas(2019)]{suvorov19} Suvorov, A.~G. \& Kokkotas, K.~D.\ 2019, \mnras, 488, 5887

\bibitem[Thompson \& Duncan(2001)]{thompson01} Thompson, C. \& Duncan, R.~C.\ 2001, \apj, 561, 980

\bibitem[Thompson et al.(2017)]{thompson17} Thompson, C., Yang, H., \& Ortiz, N.\ 2017, \apj, 841, 54

\bibitem[Vigan{\`o} et al.(2012)]{vigano12} Vigan{\`o}, D., Pons, J.~A., \& Miralles, J.~A.\ 2012, Computer Physics Communications, 183, 2042

\bibitem[Vigan{\`o} et al.(2013)]{vigano13} Vigan{\`o}, D., Rea, N., Pons, J.~A., et al.\ 2013, \mnras, 434, 123

\bibitem[Wadiasingh et al.(2020)]{wadiasingh20} Wadiasingh, Z., Beniamini, P., Timokhin, A., et al.\ 2020, \apj, 891, 82

\bibitem[Yuan et al.(2020)]{yuan20} Yuan, Y., Beloborodov, A.~M., Chen, A.~Y., et al.\ 2020, arXiv:2006.04649

\end{thebibliography}

\end{document}